\documentclass[12pt,a4paper]{article}
\usepackage{amsmath}
\usepackage{graphicx}
\usepackage{wrapfig}
\usepackage{cite}
\usepackage{amsfonts}
\usepackage{amssymb}
\begin{document}
\hfill\hbox{JINR E2-2006-119}

\hfill\hbox{PTA/06-18}

\hfill\hbox{August 2006}

\begin{center}
{\Large \textbf{Decay of Metastable Vacuum}}

\vspace{0.3cm}

{\Large \textbf{in Liouville Gravity}}

\vspace{0.5cm}

{\large A.Zamolodchikov}

\vspace{0.2cm}

Department of Physics and Astronomy, Rutgers University,

P.O.Box 849, Piscataway, New Jersey 08855-0849, USA

\vspace{0.0cm}

and

\vspace{0.0cm}

L.D.Landau Institute for Theoretical Physics RAS,

142432 Chernogolovka, Russia

\vspace{0.2cm}

and

\vspace{0.2cm}

{\large Al.Zamolodchikov}\footnote{On leave of absence from Institute of
Theoretical and Experimental Physics, B.Cheremushkinskaya 25, 117259 Moscow, Russia.}

\vspace{0.2cm}

Laboratoire de Physique Th\'eorique et
Astroparticules\footnote{UMR-5207 CNRS-UM2}

Universit\'e Montpellier II,
Pl.E.Bataillon, 34095 Montpellier, France

\vspace{0.0cm}

and

\vspace{0.0cm}

Service de Physique Th\'eorique, CNRS - URA 2306, C.E.A. - Saclay

F-91191, Gif-sur-Yvette, France
\end{center}

\vspace{1cm}

\textbf{Abstract}

A decay of weakly metastable phase coupled to two-dimensional Liouville
gravity is considered in the semiclassical approximation. The process is
governed by the ``critical swelling'', where the droplet fluctuation favors a
gravitational inflation inside the region of lower energy phase. This
geometrical effect modifies the standard exponential suppression of the decay
rate, substituting it with a power one, with the exponent becoming very large
in the semiclassical regime. This result is compared with the power-like
behavior of the discontinuity in the specific energy of the dynamical lattice
Ising model. The last problem is far from being semiclassical, and the
corresponding exponent was found to be $3/2$. This exponent is expected to
govern any gravitational decay into a vacuum without massless excitations. We
conjecture also an exact relation between the exponent in this power-law
suppression and the central charge of the stable phase.

\vspace{0.2cm}

\textbf{Preliminaries.} Nucleation mechanism in the first order phase
transitions is relatively well understood. Nucleation is the main channel of a
metastable phase decay in the case of weak metastability (see e.g.,
\cite{LL5}). The decay rate turns out to be \emph{exponentially} small in the
energy gap between the metastable and stable phases. The leading exponential
estimate is essentially simple, controlled by the energy of the ``critical
droplet''. It will be schematically repeated (in the particular case of
two-dimensional space) in the next section. More sophisticated calculation
allows for evaluation of the pre-exponential factor as well
\cite{Langer2}. Somewhat later similar mechanism has been studied
in \cite{KOV} (and then independently in \cite{Coleman}) in the context
of quantum field theory, the thermal fluctuations responsible for
the formation of critical droplet being substituted now by the
quantum tunneling. The process is commonly treated dynamically
in this case, the role of the critical droplet being played by the
``materialization of a bubble of true vacuum'' \cite{Coleman}. In the
euclidean version the critical droplet is a kind of ``instanton'', in the
sense that it is a localized solution of the classical equations of motion
(being however an unstable extremum of the action).

\begin{wrapfigure}{l}{250pt}
\includegraphics[height=3.00in, width=3.00in]
{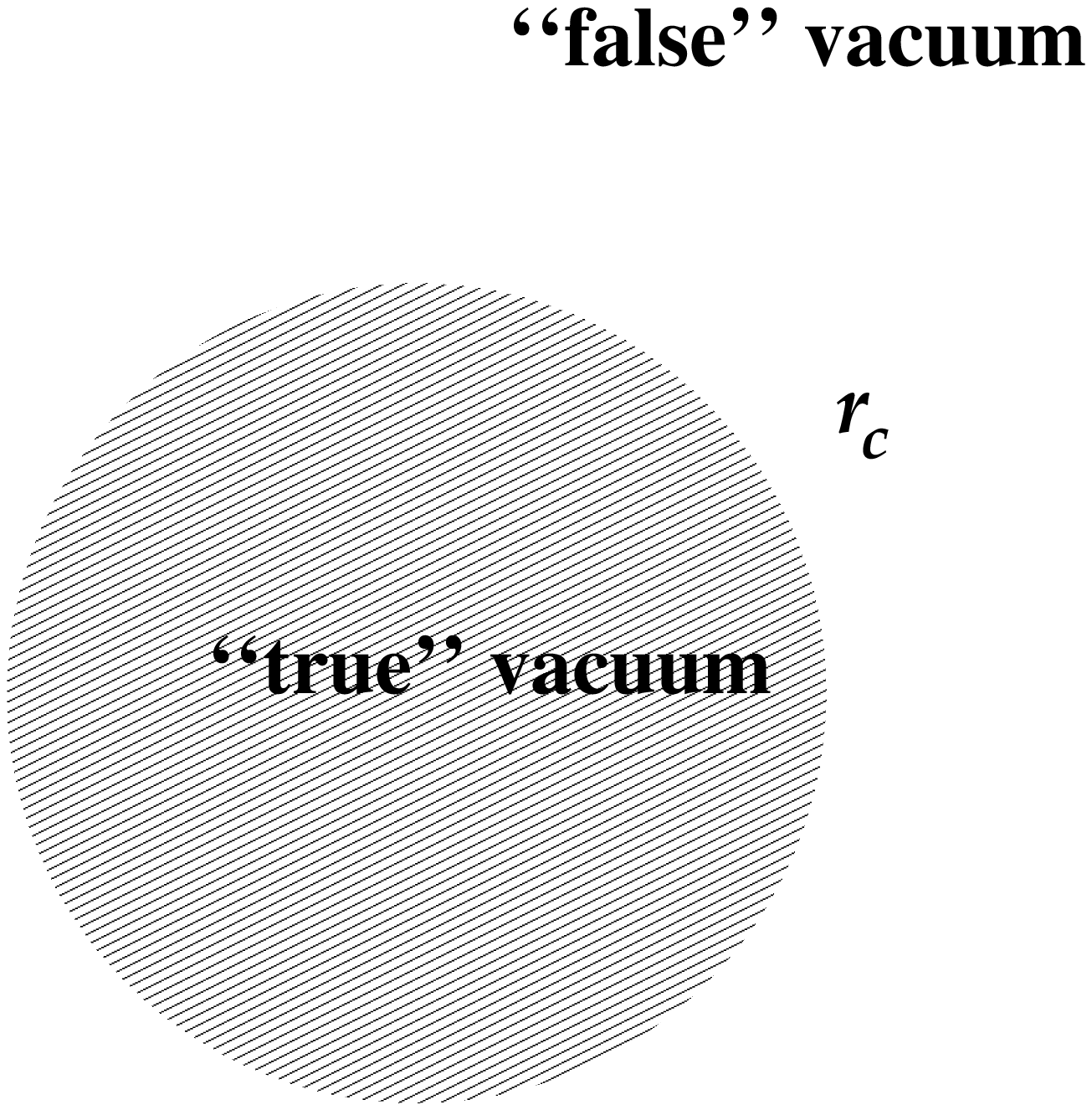}
\caption{A critical droplet fluctuation inside the spatially
infinite metastable phase}
\label{droplet}
\end{wrapfigure}
Another facet of this phenomenon can be seen when looking at the specific free
energy as an analytic function of the energy gap $\mu$ between the phases
(e.g., $\mu$ is proportional to the magnetic field strength $h$ in the case of
low temperature phase ferromagnet in a weak field). When continued to the
values corresponding to the metastable state, the free energy exhibits a
branch cut with pure imaginary discontinuity, directly traceable to the
``critical droplet'' fluctuations (in particular, the discontinuity is
exponentially small in the absolute value of $\mu$, as in the Eq.\eqref{Pexp}
below) \cite{Andreev, Fisher, Langer1}. It is essential that the singularity
appears only in the thermodynamic limit, when the size of the system is sent
to infinity. In the field-theoretic context the above specific free energy is
interpreted as the vacuum energy density, and the presence of the imaginary
part simply means that the ``false vacuum'' is the quantum-mechanically
metastable state, thus suggesting direct interpretation of the above
discontinuity in terms of the decay probability \cite{KOV, Coleman}. In the
context of the thermal phase transition, a close qualitative correlation
between the discontinuity of the free energy and the rate of decay of the
metastable phase is widely assumed, while exact relation has been established
in the limit of weak metastability only \cite{Langer2}.

The subject attracts much attention. Whereas the stat-mechanical aspects are
of obvious interest in a wide range of disciplines from chemistry and
metallurgy to astrophysics, the interest in the ``false vacuum'' decay is
mainly due to the natural concern about the fate of our Universe, if it faces
the vacuum metastability disaster \cite{shit}\footnote{sometimes seen as ``the
ultimate ecological catastrophe'', see \cite{CL} for the developed concept.}.
There are even more worries after Coleman and De Luccia had studied the
effects of gravity on this decay \cite{CL}. Although the
Einstein gravity hardly influences the decay rate, the ``realistic''
parameters taken into account, the vacuum formed ``a microsecond after''
hardly can comfort human beings. For more similar horrors and related ethical
problems see \cite{ethic}, and if even that is not enough, also
\cite{Agor}. Another interesting, although less dramatic, are the implications
in the cosmology, the scenario where the whole drama is well in the past
\cite{shit2}.

Turn to less apocalyptic matters. It has been observed recently that the
effect of quantum gravity in two dimensions changes qualitatively the
character of the discontinuity in the specific free energy \cite{term}. A
lattice version of quantum gravity coupled to Ising spins, known as the
dynamical lattice Ising model, is exactly solvable \cite{BK} through the
matrix model technique. The vacuum energy as a function of the temperature and
magnetic field allows analytic expression. In the case of ordered phase, at
small absolute values of the magnetic field there is indeed a branch cut, the
discontinuity being suppressed as a \emph{power} rather then exponentially.

\begin{wrapfigure}{r}{230pt}
\includegraphics[height=4.00in, width=3.00in]
{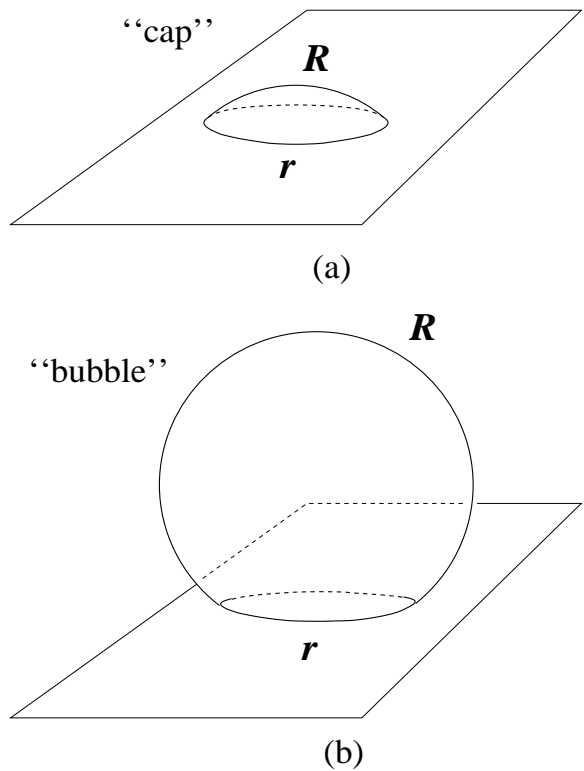}
\caption{``Critical swellings in the case $s<1$ (a) and $s>1$ (b).}
\label{capbubble}
\end{wrapfigure}
Some details of this observation are briefly recapitulated in one of the
subsequent sections. It suggests naturally that this power law also governs
the decay rate in the limit of weak metastability. We find it therefore
instructive to repeat the semiclassical critical droplet calculation taking
into account relevant effects of the quantum Liouville gravity. The latter,
rather then the Einstein gravity, governs the dynamics of the geometry in two
dimensions \cite{Polyakov}. The Liouville gravity is characterized by a
dimensionless coupling constant, related to the central charge $c$ of the
massless modes around the vacuum. The semiclassical limit corresponds to large
negative $c\rightarrow-\infty$. In this limit we find that the decay is
controlled by a gravitational version of critical droplet, which we call the
``critical swelling''. This localized distortion of the metric is induced by
the nucleus of the ``true'' vacuum. It looks more like a ``cap'' at the
extreme semiclassical limit $c\mu\rightarrow-\infty$ and like a ``bubble'' in
the opposite weak metastability limit $c\mu\rightarrow-0$. In the last case
the decay rate behaves as $\mu^{-c/6}$ at $\mu\rightarrow0$, the
\textit{power} law replacing the exponential suppression in the flat.

This result is semiclassical. It is natural to expect in general some
\emph{power} behavior
\begin{equation}
P\sim\mu^{\delta}\,, \label{mudel}%
\end{equation}
with the parameter $\delta$ depending on the central charge $c$ of the stable
vacuum. An interesting question arises about the exact dependence of this
exponent on $c$. From the exact solution one can figure out that $\delta=3/2$
for the decay into a vacuum without massless modes, where $c=0$. At the end of
this paper we develop arguments in favor of the following exact formula
\begin{equation}
\delta=\frac{13-c+\sqrt{(1-c)(25-c)}}{12} \label{ac}%
\end{equation}
which gives $3/2$ at $c=0$ while $\delta\sim(13-c)/6+O(1/c)$ in the
semiclassical limit.

\textbf{Nucleation in flat background.} To begin with we recall the standard
consideration about the decay of a metastable state in statistical mechanics,
following \cite{LL5}, \textsection162. Formally identical analysis applies to
the semiclassical tunneling decay of the ``false vacuum''. Here we restrict
the discussion to the leading exponential factor, although more accurate
treatment, which allows to estimate the pre-exponential multiplier too, can be
carried out, see \cite{KOV, Coleman}. In \cite{Volosha} the two-dimensional
version is specially treated in very details.

Consider a flat infinite two-dimensional plane filled up with a metastable
vacuum\footnote{Considerations of this paragraph go in a space of any
dimensionality as well. Here we take it two-dimensional to match with what
follows.}. Its specific energy $\mathcal{E}_{1}$ is meant to be slightly above
the energy of the ``true'' stable vacuum $\mathcal{E}_{0}$. We denote $\mu$
the (positive) difference between the two energies, i.e., $\mathcal{E}%
_{1}=\mathcal{E}_{0}+\mu$. Imagine now a fluctuation inside this false sea,
where a region $\Gamma$ of true vacuum is formed, and let $A(\Gamma)$ and
$p(\Gamma)$ be respectively the area and perimeter of $\Gamma$. Also, let
$\sigma>0$ be the surface tension at the border between true and the false
vacuum. The energy of the fluctuation $\Delta E(\Gamma)=p(\Gamma
)\sigma-A(\Gamma)\mu$ is a result of a competition between the gain in the
bulk and the expense at the perimeter. Weak metastability means that the
characteristic $\Delta E$ is much smaller then the temperature (in the field
theoretic context $\Delta E$ is interpreted as the action of the ``instanton''
fluctuation and is compared with the Plank constant). In this case the
estimate for the probability $P(\Gamma)$ of this fluctuation is
\begin{equation}
P(\Gamma)\sim\exp\left(  -\frac{\Delta E(\Gamma)}{T\;(\hbar)}\right)
\label{PDE}%
\end{equation}
where $T$ (or the Plank's constant $\hbar$ in the case of quantum tunneling)
is the parameter governing the fluctuations. In what follows we always
imply that the energy (action) is measured in the units of $T$ ($\hbar$)
and omit the corresponding factors.

By usual arguments \cite{KOV, Coleman}, the specific (per unit area of the 2D
space) probability of the false vacuum decay is controlled mainly by the
configuration which extremizes (\ref{PDE}), the decay instanton, a.k.a. the
critical droplet. The instanton is a circle of radius
\begin{equation}
r_{\text{c}}=\sigma/\mu\label{rc}%
\end{equation}
(fig.\ref{droplet}). In this calculation we assume that the size of the
critical droplet is large as compared to all microscopic scales, such as the
thickness of the border of the droplet etc. This is certainly the case in the
weak metastability limit $\mu\rightarrow0$ (as the eq.(\ref{rc}) indicates).
Evaluating the energy of the critical droplet one finds
$\Delta E_{\text{c}}=\pi\sigma^{2}/\mu$.
This means that at small $\mu$ the decay rate is exponentially suppressed
\begin{equation}
P\sim\exp\left(-\frac{\pi\sigma^{2}}{\mu}\right)  \label{Pexp}%
\end{equation}
in the energy gap $\mu$.

\textbf{Nucleation in 2D Liouville gravity. }Now we're going to couple our
system to the gravitating background, where the effective action of the
gravity is given by the Liouville action $A_{\text{L}}$ \cite{Polyakov}.
Choosing the coordinate system where $g_{ab}=e^{\varphi}\delta_{ab}$ we have
\begin{equation}
A_{\text{L}}=\frac k{96\pi}\int\left(  \partial_{a}\varphi\right)  ^{2}d^{2}x
\label{AL}%
\end{equation}
Parameter $k$ is related to the effective number of massless modes in the
corresponding vacuum, called the central charge $c$ \cite{BPZ}. In the
semiclassical limit $c\rightarrow-\infty$ it has been found that $k\sim-c$
\cite{Polyakov}. In general the relation between $k$ and $c$ depends on the
regularization prescription. Regularization with respect to a fixed background
metric implies the following expression \cite{DDK}
\begin{equation}
k=6b^{-2} \label{kc}%
\end{equation}
in terms of the standard Liouville parameter $b$ related to $c$ through
\begin{equation}
26-c=1+6(b^{-1}+b)^{2} \label{bc}%
\end{equation}

As in the case of the flat background, here we are interested in the specific
(per unit area) probability of the decay in the limit when the area is very
large \footnote{This limit can be achieved by fine-tuning the ``bare''
cosmological constant $\lambda_{0}$ to a certain critical value $\lambda_{c}$,
so that the physical cosmological constant vanishes. That is how it is done in
the matrix models of gravity (see e.g. example in the next section). This
critical value (more precisely, $-\lambda_{c}$) is interpreted as
the intensive part of the specific free energy, and it is this quantity which
develops the singularity $\mu^{\delta}$ at $\mu\to0$.}. Correspondingly, our
setting of the decay problem is as follows: at the beginning we have an
infinite, globally flat space filled with the metastable phase.
Semiclassically this is described by the solution $\varphi=$ const, which we
choose to be $\varphi=0$. Imagine a local fluctuation somewhere near the
origin, so that $\varphi=0$ for $\left|  x\right|  >r$ with some $r$. At
$\left|  x\right|  <r$ we have a seed of the stable phase, which has negative
energy density $-\mu$ with respect to that of the false one. Inside the
droplet we are looking for the extremum of the functional
\begin{equation}
A_{\text{L}}=\int_{\left|  x\right|  <r}\left(  \frac k{96\pi}\left(
\partial_{a}\varphi\right)  ^{2}-\mu e^{\varphi}\right)  d^{2}x \label{ALmu}%
\end{equation}
with respect to the scale function $\varphi(x)$ at $\left|  x\right|  <r$. The
extremum is of course a solution to the positive curvature Liouville equation
\begin{equation}
\partial\bar\partial\varphi=-12\pi\mu k^{-1}e^{\varphi} \label{Leq}%
\end{equation}
The general solution respecting the symmetry of the problem is a sphere
\begin{equation}
e^{\varphi}=\frac{4R^{2}a^{2}}{\left(  1+a^{2}x\bar x\right)  ^{2}}
\label{sphere}%
\end{equation}
of radius $R=R_{\text{c}}$ with
\begin{equation}
R_{\text{c}}^{2}=\frac k{24\pi\mu} \label{Rc}%
\end{equation}
and a scale parameter $a$ to be determined from the continuity at the boundary
of the swelling. It will turn instructive not to imply the relation
$R=R_{\text{c}}$ from the very beginning considering $\mu$ and $R$ as
unrelated parameters, to be bound later in the extremum condition.

We need to sew this solution with the outside region in a continuous way,
i.e., at $\left|  x\right|  =r$ we have $2Ra=1+a^{2}r^{2}$. This quadric has
two solutions for $a$
\begin{equation}
a_{\pm}=\frac{R\pm\sqrt{R^{2}-r^{2}}}{r^{2}} \label{apm}%
\end{equation}
Here $a_{+}$ corresponds to the ``bubble'' solution, while $a_{-}$ is a
``cap'' one, see fig.\ref{capbubble}. Both are conveniently uniformized
through the parameterization
\begin{equation}
2R=r(t+t^{-1}) \label{Rrt}%
\end{equation}
In this way both branches are covered if we set $ar=t$ and let $t$ run from
$0$ to $\infty$, $t<1$ corresponding to the ``cap'' configuration and
$t>1$ being related to the ``bubble''. It is easy to evaluate in this
parameterization both terms of the Liouville action (\ref{ALmu})
\begin{align}
\frac k{24\pi}\int\partial\varphi\bar\partial\varphi d^{2}x  &  =\frac
k6\left(  \log(1+t^{2})-\frac{t^{2}}{1+t^{2}}\right) \label{Lterms}\\
\mu\int_{\left|  x\right|  <r} e^{\varphi}d^{2}x  &  =4\pi\mu R^{2}\frac{t^{2}}{1+t^{2}}\nonumber
\end{align}
Adding the surface tension term $2\pi\sigma r$ we find the total energy of the
swelling to be varied
\begin{equation}
E=\frac k6\left(  \log\left(  1+t^{2}\right)  -\frac{t^{2}}{1+t^{2}}%
-x^{2}(1+t^{2})+2sx\right)  \label{Etx}%
\end{equation}
were we have used the notation
\begin{equation}
s=\sigma\sqrt{\frac{6\pi}{k\mu}} \label{s}%
\end{equation}
instead of the surface tension. Also we introduced another parameter
\begin{equation}
2x=r/R_{\text{c}}\ . \label{x}%
\end{equation}

\begin{wrapfigure}{r}{240pt}
\includegraphics[height=3.00in, width=3.00in]
{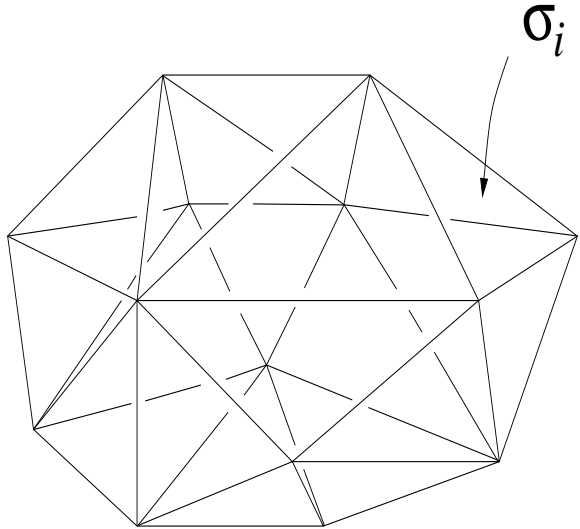}
\caption
{An example of triangulation of a sphere $g=0$. Ising spins are attached to the
triangles.}
\label{spinsw}
\end{wrapfigure}
The relevant extremum is achieved at $t=s$ and $x=s/(1+s^{2})$. The extremal
value
\begin{equation}
E_{\text{c}}=\frac k6\log\left(  1+s^{2}\right)  \label{Ec}%
\end{equation}
It can be verified directly that in both cases the chosen extremum corresponds
to a saddle point in the parameters $(R,r)$, in the case $s\ll 1$ the unstable
direction being mostly along the ``perimeter'' direction $r$, while in the
bubble case $s>1$ it is generally along the ``inflation'' direction $R$.
Finally, we find for the decay rate
\begin{equation}
P\sim\left(  1+\frac{6\pi\sigma^{2}}{k\mu}\right)  ^{-k/6} \label{Ppow}%
\end{equation}
At the extreme limit $k\rightarrow\infty$ we are back to the classic droplet
result (\ref{Pexp}). On the contrary, the weak metastability limit
$\mu\rightarrow0$ is characterized by a power-like suppression $P\sim\left(
k\mu/\sigma^{2}\right)  ^{k/6}$.

\textbf{Gravitational Ising vacuum energy.} We recall in this section very
briefly the dynamical lattice Ising model (DLIM), as it has been proposed in
\cite{BK}, and present without derivation the form of its exact solution in
the scaling limit near the ferromagnetic phase transition. The formulation is
standard for the two dimensional random lattice models and starts with an
ensemble $\left\{  G_{N}^{(g)}\right\}  $ of planar graphs (in general of
genus $g$) and of size $N$. It is considered well established that the scaling
singularity, which is supposed to correspond to a continuous limit, doesn't
depend on the details of this ensemble (neither on most microscopic details of
the spin model located on the graph). Here for definiteness we will imply a
very particular realization of $\left\{  G_{N}^{(g)}\right\}  $ as the graphs
made exclusively of triangles, the triangulations, like an example in
fig.\ref{spinsw}. In this case the size $N$ can be given precise meaning, the
number of triangles in $G_{N}^{(g)}$.

In DLIM we associate a spin variable $\sigma_{i}=\pm1$ with each triangle
$i=1,2,\ldots,N$. As in the usual Ising, only ``nearest neighbor'' triangles
$\left\langle ij\right\rangle $ (i.e., having a common edge) contribute to
spin-spin interaction
\begin{equation}
\mathcal{H}\left[  \{\sigma\}\right]  =\sum\nolimits_{\left\langle
ij\right\rangle }K\sigma_{i}\sigma_{j}+\sum\nolimits_{i}H\sigma_{i}
\label{Ising}%
\end{equation}
where $K$ is the ``thermal'' parameter (called traditionally the exchange
integral) and $H$ is the external magnetic field. In this study we
consider the large $N$ limit characteristics, namely the specific free energy
(per triangle) of DLIM. This energy is an intensive characteristic and as such
doesn't depend of global properties, like $g$. For our purposes it is most
convenient to start with the spherical partition function ($g=0$), where the
result of \cite{BK} applies directly. Introduce the microcanonic spherical
partition function
\begin{equation}
Z_{N}(K,H)=\sum\nolimits_{\left\{  G_{N}^{(0)}\right\}  }\sum
\nolimits_{\{\sigma\}}\exp\left(  -\mathcal{H}\left[  \{\sigma
\}\right]  \right)  \label{ZN}%
\end{equation}
Here, in addition to the standard Ising sum over the spin configurations
$\{\sigma\}$ the sum over $\left\{  G_{N}^{(0)}\right\}  $ is also
implied, the last being the lattice version of the quantum gravity path
integral over the metrics. Considering the leading logarithmic behavior at
large $N$%
\begin{equation}
\log Z_{N}(K,H)\sim-N\mathcal{E}(K,H) \label{EKH}%
\end{equation}
we readily see that the intensive energy $\mathcal{E}(K,H)$ is determined by
the position of the rightmost singularity in the grand partition function
\begin{equation}
Z(K,H,X)=\sum\nolimits_{N}e^{-XN}Z_{N}(K,H) \label{Zgrand}%
\end{equation}
in the ``chemical potential'' $X$.

Let $(X,K,H)=(X_{\text{c}},K_{\text{c}},0)$ be the position of the double
singularity in the grand partition function, caused by the combined divergence
in size $N$ in (\ref{Zgrand}) and of (\ref{ZN}) due to the long correlations
of spins at the ferromagnetic phase transition. The singular part
$Z_{\text{sing}}(x,t,h)$ of the partition function depends in a scaling way on
the deviations $x\sim X-X_{\text{c}}$, $t\sim K_{\text{c}}-K$ and $h\sim H$
(precise definitions depend on the choice of scale). Boulatov and Kazakov
\cite{BK} showed that the function
\begin{equation}
u=\partial^{2}Z_{\text{sing}}/\partial x^{2} \label{u}%
\end{equation}
solves the following simple algebraic equation
\begin{equation}
x=u^{3}+\frac{3}{2}tu^{2}+\frac{h^{2}}{(u+t)^{2}} \label{BKeq}%
\end{equation}
As usual the continuous theory deals with the \emph{singular} part of the
intensive energy, which, up to sign coincides with $x_{\text{c}}$, the
singularity of equation (\ref{BKeq}). It is easy to see \cite{term} that the
relevant singularity is located at
\begin{equation}
x_{\text{c}}=\frac{t^{3}}{2}f\left(  5f^{2}+9f+3\right)  \label{xc}%
\end{equation}
where $f$ is an appropriate solution to the following fifth order algebraic
equation
\begin{equation}
f(f+1)^{4}=\frac{2h^{2}}{3t^{5}} \label{fifth}%
\end{equation}
As it is argued in ref.\cite{term}, the low temperature phase $t<0$ of
DLIM corresponds to the solution
\begin{equation}
f=-1+\sum_{n=1}^{\infty}\frac{\Gamma(5n/4-1)}{n!\Gamma(n/4)}\left(
\frac{2^{1/4}h^{1/2}}{3^{1/4}(-t)^{5/4}}\right)  ^{n} \label{f}%
\end{equation}
and leads to the following singularity at $h\rightarrow0$%
\begin{align}
t^{-3}x_{\text{c}}  &  =\frac{1}{2}+3\sum_{n=2}^{\infty}\frac{(n-1)\Gamma
(5n/4-3)}{n!\Gamma(n/4)}\left(  \frac{2^{1/4}h^{1/2}}{3^{1/4}(-t)^{5/4}%
}\right)  ^{n}\label{xexpand}\\
&  =\frac{1}{2}-\frac{\sqrt{6}h}{(-t)^{5/2}}+ \frac{2^{3/4}h^{3/2}}%
{3^{3/4}(-t)^{15/4}}+\frac{h^{2}}{4(-t)^{5}}+\ldots\nonumber
\end{align}
The free energy exhibits the square root branch cut at $h=0$ with the exponent
$3/2$.

\textbf{Discussion and Speculations.} There are serious reasons to believe
that in two dimensions the Liouville gravity affects the standard nucleation
mechanism of the metastable decay in a significant way. The standard
exponential suppression of the decay rate is substituted by a weaker
power-like one, changing qualitatively the picture. We were only able to
obtain more precise information either in the semiclassical limit, controlled
by the instanton, or in the exactly solvable dynamical lattice Ising model.
One could expect in principle a one-parameter family of universality classes,
depending on the central charge in the final state of the decay. A better
understanding of this family is desired. In this relation different steps
would be valuable. One thing, which is not attempted in the present study, is
the one-loop (and higher) corrections to the critical swelling instanton. Such
calculation would provide a constant term in the semiclassical expansion
$\delta\simeq-c/6+\mathrm{{const}}$ of the exponent in \eqref{mudel}, and thus
support (or disprove) the conjectured exact form \eqref{ac}. It is also useful
to look for more solvable examples among the dynamical lattice statistical
systems exhibiting metastability and, preferably, having non-trivial massless
modes in the stable vacuum. And of course, metastability opens a large
playground for Monte-Carlo simulations, essentially dynamical processes. Many
appropriate models are known (e.g., the ``generalized random lattice Ising
model of ref.\cite{Ishimoto}), which certainly can be put to metastable state,
while the central charge in the stable vacuum remains completely under our
control.
\begin{wrapfigure}{l}{250pt}
\includegraphics[height=1.80in, width=3.00in]
{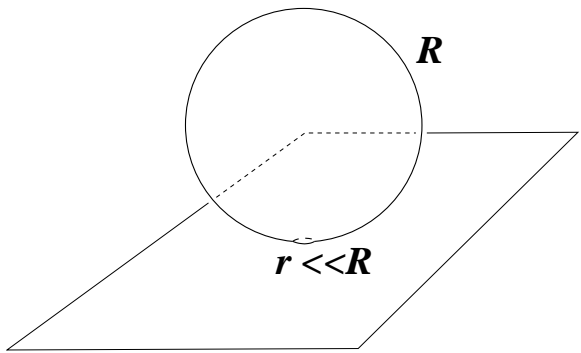}
\caption{Critical swelling in the regime of extremely weak metastability $\mu
\to0$.}
\label{puncture}
\end{wrapfigure}

The Eq.\eqref{mudel} applies to the decay probability in the situation when
the metastable phase is accommodated by infinite globally flat geometry, or
more generally when all the length scales associated with that geometry are
large as compared with $\mu^{-1}$. In particular, the physical cosmological
constant of the metastable phase must be much smaller then $\mu$. This
settlement is perfectly appropriate for the situation of the previous
sections, where we were interested in the intensive part of the specific free
energy. But the semiclassical calculations of the third section certainly can
be generalized to incorporate the effect of the cosmological constant in an
interesting way.

To conclude, we speculate a conjecture for the exact relation (\ref{ac})
between the central charge and the exponent in the decay law. It is easy to
see, that in the limit of extremely weak metastability $\mu\rightarrow0$ the
``bubble'' configuration of fig. \ref{capbubble}(b) tends to an almost
complete sphere, connected to the metastable plane through a small
``throat''. The last can be seen as a pointlike puncture (fig.\ref{puncture}).
We recall in this context that the partition function of a compact sphere
behaves as $\mu^{b^{-2}+1}$ where $b$ is from (\ref{bc}). This is an exact
result of quantum Liouville field theory, which can be found e.g., in
\cite{KPZ}. A puncture corresponds to a differentiation with respect to $\mu$
and the action of the limiting swelling is evaluated as $\mu^{1/b^{2}}$, i.e.,
the power law conjectured in (\ref{ac}). In this consideration we assume (with
no really good reasons) that the essentially semiclassical arguments about the
critical swelling, which we used to reduce the problem to a punctured sphere,
apply qualitatively also at general quantum $c$.

Let us mention also that the exponent equal to $1/b^2+1$ (which differs by
$1$ from \eqref{ac}) was proposed recently by A.Polyakov
\cite{Polyakov2} in apparently different context, in relation
with possible unstability of the de Sitter space. Obviously, understanding
precise relation to the decay problem considered here would be of much
interest.

\textbf{Acknowledgments}

A.Z is grateful to A.Sorin and the Bogoliubov Laboratory of Theoretical
Physics in Dubna for warm hospitality extended to him during the summer of
2006. Research of AZ is supported in part by the DOE grant \# DE-FG02-96 ER 40949

Al.Z thanks the organizers of the international School-Workshop ``Calculations
for future colliders'', July 15--25, for the stimulating scientific
atmosphere, which played important role in finishing this study. He was also
supported by the European Committee under contract EUCLID HRPN-CT-2002-00325
and by the INTAS project under grant \#INTAS-OPEN-03-51-3350. Discussions with
I.Kostov at the early stages of the work were quite stimulating.

\end{document}